\documentstyle[12pt]{article}
\begin{document}
\begin{flushright}
Alberta-Thy-43-96\\
November, 1996\\
\end{flushright}

\begin{center}
{\Large \bf  Direct CP  Violation  Asymmetries in  Exclusive  B Decays in a Bethe-Salpeter Approach}\\
\vskip 5mm
A.N.Kamal and C.W.Luo\\
{\em Theoretical Physics Institute and Department of Physics,\\ University of Alberta,
Edmonton, Alberta T6G 2J1, Canada.}\\
\end{center}

\vskip 5mm
\begin{abstract}
 CP asymmetry in some exclusive decays of charged B  meson are  calculated in a Bethe-Salpeter   approach.  Hadronic final state interactions are ignored. Complex decay amplitudes are assumed to arise entirely from perturbative quark-antiquark loops. Calculations are done both with and without the gluon quark-antiquark vacuum polarization loops. The effects of neglecting the imaginary parts arising from the diagonal quark-antiquark loops are also studied.  
\end{abstract}

\newpage

\begin{flushleft}
{\Large \bf 1. Introduction}
\end{flushleft}

It has long been recognised that CP asymmetry in partial decay rates can be generated in processes that  involve (i) at least two different Cabibbo-Kobayashi-Maskawa (CKM) angles and (ii) at least two different strong interaction phases\cite{KPS}.  These strong phases could be generated either perturbatively through the penguin processes or  through hadronic final state interactions (fsi). In processes involving a single  isospin in the final state, CP asymmetry is still possible provided that the penguin processes generate strong phases\cite{DHP}. \\

\par
The commonly followed practice in evaluating decay matrix elements has been  to use the factorization approximation which reduces the computation to the knowledge of certain form factors. This method treats $ q^2 $, the invariant momentum transfer carried by the gluon or the photon in the strong or electroweak penguins, as a free parameter whose value is usually chosen in the range $ m_b^2/4\leq q^2\leq m_b^2/2 $ for B decays. In B decays,  perturbative strong phases generated by $c\bar{c} $ states in the penguin loops are particularly sensitive to the choice of $q^2 $. In contrast to the factorization scheme where one is forced to choose an effective value of $ q^2 $, in a complete calculation  $ q^2$ would be  integrated over and would not remain a free parameter. \\ 

In the past, Simma et.al.\cite{SW} have employed the formalism of Lepage and Brodsky\cite{LB}
to calculate CP asymmetries in inclusive and exclusive decays of charged B meson. In their approach, due to the fact that extra gluons have to be used to accelerate the spectator quark, new  absorptive parts are generated. They also found that absolute  partial rates were too low by as much as two orders of magnitude. They had to scale up the rates so that the semileptonic rate $B\rightarrow Dl\bar{\nu} $ was correctly normalized. It is our view, as shown later,  that it is doubtful if a single process-independent scale factor would be adequate for all exclusive channels. The advantage of their method, aside from pedagogy, is that it goes beyond the factorization approximation and, consequently, $q^2 $ is not left as a free parameter. \\

In our work, we have used the Bethe-Salpeter (B-S)  method of Mandelstam\cite{SM} and Nishijima\cite{KN} to treat the bound state mesons. This method has been applied previously to other decay processes, including radiative decay\cite{E}. Before embarking on the discussion of our calculations, we wish to state that our results on  CP asymmetries are somewhat different in size, though not in sign, to those of \cite{SW}. This is not unexpected considering the differences in  methodology and approximations used. The results, as emphasized in\cite{SW}, are trustworthy to within a factor of 2. However, we share with\cite{SW}  the diffculty in getting the absolute rates right.  We also need to normalize the calculated rates such that the semileptonic rates are correctly normalized.  This, however, does not affect CP asymmetry as absolute normalizations cancel in the definition. \\

\begin{flushleft}
{\Large \bf  2. Preliminaries}
\end{flushleft}

We work with the following effective Hamiltonian for $b\rightarrow s $ transition (for $b\rightarrow d $ transition, replace s by d)\cite{DH}

\begin{equation}
H_{eff}={G_F  \over \sqrt{2} } \sum_{q=u,c}{\left\{  V_{qb} V^*_{qs}[ C_1 O^q_1+C_2 O^q_2+
\sum_{i=3}^{10}{C_i O_i}]  \right\} } 
\end{equation}
where $V_{qb} $ etc are the CKM matrix elements, $ C_i $ are the Wilson coefficients (discussed below) and $O_i $ are the following $Current\otimes Current $ operators:

\par
Tree level operators:
\begin{equation}
O^q_1=(\bar{s}q)_{V-A} (\bar{q}b)_{V-A}, ~~~~~O^q_2=(\bar{s}_{\alpha}q_{\beta})_{V-A} (\bar{q}_{\beta} b_{\alpha})_{V-A};
\end{equation}

QCD penguin operators:
\begin{eqnarray}
O_3=(\bar{s}b)_{V-A}\sum_{q^{\prime}} (\bar{q}^{\prime}q^{\prime})_{V-A},  & 
O_4=(\bar{s}_{\alpha} b_{\beta})_{V-A}\sum_{q^{\prime}} (\bar{q}^{\prime}_{\beta} q^{\prime}_{\alpha})_{V-A},     \nonumber  \\
O_5=(\bar{s}b)_{V-A}\sum_{q^{\prime}} {(\bar{q}^{\prime}q^{\prime})_{V+A}}, & 
O_6=(\bar{s}_{\alpha} b_{\beta})_{V-A}\sum_{q^{\prime}}{(\bar{q}^{\prime}_{\beta} q^{\prime}_{\alpha})_{V+A}};
\end{eqnarray}

 Electroweak penguin operators:
\begin{eqnarray}
O_7={3 \over 2}(\bar{s}b)_{V-A}\sum_{q^{\prime}} {(e_{q^{\prime}}\bar{q}^{\prime}q^{\prime})_{V+A}}, & 
O_8={3\over 2}(\bar{s}_{\alpha} b_{\beta})_{V-A} \sum_{q^{\prime}}{(e_{q^{\prime}}\bar{q}^{\prime}_{\beta} q^{\prime}_{\alpha})_{V+A}},
\nonumber \\
O_9={3 \over 2}(\bar{s}b)_{V-A}\sum_{q^{\prime}} {(e_{q^{\prime}}\bar{q}^{\prime}q^{\prime})_{V-A}}, & 
O_{10}={3\over 2}(\bar{s}_{\alpha} b_{\beta})_{V-A} \sum_{q^{\prime}}{(e_{q^{\prime}}\bar{q}^{\prime}_{\beta} q^{\prime}_{\alpha})_{V-A}}
\end{eqnarray}
where the subscripts $V\pm A $ represent $\gamma_{\mu} (1\pm \gamma_5)$, $\alpha$ and $\beta$ are color indices.  $\sum_{q^{\prime}}$ 
represents a sum over u,d,s and c quarks.
\par
Wilson coefficients, $C_i$,  are renormalization scheme dependent when the next -to-leading order corrections are considered. However, as the physical quantities should be scheme-independent, one needs to define effective Wilson coefficients  to be used in computation. The manner in which these effective coefficients are defined is detailed in \cite{DH, RF, BJLW}. We only summarize the final result: The effective coefficients are 
\begin{eqnarray}
C^{eff}_1=\bar{C}_1,~~  C^{eff}_2=\bar{C}_2, & ~~ C^{eff}_3=\bar{C}_3 - P_s / N_c,  & C^{eff}_4=\bar{C}_4 + P_s,  \nonumber   \\
C^{eff}_5=\bar{C}_5 - P_s/N_c,~~~~ &  C^{eff}_6=\bar{C}_6 +P_s, &  C^{eff}_7=\bar{C}_7 +P_e, 
 \nonumber   \\
 C^{eff}_8=\bar{C}_8, ~~~~~~~~~~~~~~ & C^{eff}_9=\bar{C}_9 +P_e, &  C^{eff}_{10}=\bar{C}_{10}.
\end{eqnarray}
where $N_c $ is the number of colors and 
\begin{eqnarray}
& \bar{C}_1=1.1502, ~\bar{C}_2=-0.3125, ~\bar{C}_3=0.0174, ~\bar{C}_4=-0.0373, ~\bar{C}_5=0.0104,~ \bar{C}_6=-0.0459, & \nonumber   \\
& \bar{C}_7=-1.050\times 10^{-5}, ~~\bar{C}_8=3.839\times 10^{-4}, ~~\bar{C}_9=-0.0101, 
~~\bar{C}_{10}=1.959\times 10^{-3}.~~~~~~&
\end{eqnarray}

To one-loop approximation, the contribution to  $ P_{s,e} $ from penguin diagrams with insertions of  tree operators $ O_{1, 2} $ (see Fig.1(a)),  are 

\begin{eqnarray}
P^{(1)}_s &= &{\alpha_s (\mu) \over  8\pi} C_1 (\mu) [{10 \over 9}+\frac{2}{3}\ln{\frac{m_q^2}{\mu^2}}-G(m_q, \mu, q^2)], \\
P_e &= &{\alpha_{em} (\mu) \over  3\pi} [C_2 (\mu)+{C_1(\mu) \over N_c}] [{10 \over 9}+\frac{2}{3}\log{\frac{m_q^2}{\mu^2}}-G(m_q, \mu, q^2)].
\end{eqnarray}

Here the flavor q stands for u and c in the penguin loop and $q^2 $ is the invariant  momentum carried by the gluon or the photon.  $G(m_q, \mu, q^2) $ is given by
\begin{equation}
G(m_q, \mu, q^2)= -4\int_{0}^{1}{dx x (1-x) \ln{[1-x (1-x) {q^2 \over  m^2_q}}]},
\end{equation}
Obviously,  as $q^2> 4 m^2_q$, a strong perturbative phase is generated  by virtue of  $G(m_q, \mu, q^2)$ becoming complex. This perturbative phase arising from the complex  nature of $G(m_q, \mu, q^2) $ has nothing to do with the strong phases generated by long range fsi among hadrons. The latter respect isospin symmetry,  while gluons operating in color-space, the perturbative phases have nothing to do with isospin space. 

\par
In order to accommodate some of the cancellations required by the CPT theorem\cite{GH}, we also consider a 1-loop correction to the gluon propagator, Fig.1(b), leading to a two-loop expression for $P_s$,

\begin{equation}
P^{(2)}_s={1\over 2}({\alpha_s (\mu) \over  4\pi})^2 C_1 (\mu) [{10 \over 9}+\frac{2}{3}\ln{\frac{m_q^2}{\mu^2}}-G(m_q, \mu, q^2)] \sum_{q^{\prime}}{[ 
\frac{2}{3}\ln\frac{m_{q^{\prime}}^2}{\mu^2}-G(m_{q^{\prime}}, \mu, q^2)  ]}.
\end{equation}

It is important to note that we will not concern ourselves with the soft hadronic fsi; all phases are considered to be generated by the complex value of $G(m_q, \mu, q^2) $.\\

\begin{flushleft}
{\Large \bf 3. Calculation of hadronic matrix elements}
\end{flushleft}
\par

The Feynman rules for the B-S formalism of  Mandelstam\cite{SM} and Nishijima\cite{KN}  have been abstracted in Esteve et.al.\cite{E}. We illustrate the application of those rules by writing down the matrix element for $B^-\rightarrow D^-D^0 $ transition. In Fig.2, we have shown the tree and the penguin graphs

\begin{equation}
A_{Fig.2(a)}=i\int \frac{d^4s}{(2\pi )^4}\frac{d^4t}{(2\pi )^4} T_r[\chi^{P_1}_{B}(s)(\not s-\frac{\not P_1}{2}-m_u)
\bar{\chi}^{P_2}_{D^0}(s-\frac{P_3}{2})\gamma_{\mu}(1-\gamma_5)\bar{\chi}^{P_3}_{D^-}(t)
\gamma^{\mu}(1-\gamma_5)],
\end{equation}
\begin{equation}
A_{Fig.2(b)}=i\int \frac{d^4s}{(2\pi )^4}\frac{d^4q}{(2\pi )^4} C_i(q^2) T_r[\chi^{P_1}_{B}(s)(\not s-\frac{\not P_1}{2}-m_u)
\bar{\chi}^{P_2}_{D^0}(s-\frac{P_3}{2})\Gamma_{\mu}\bar{\chi}^{P_3}_{D^-}(s-q+\frac{P_2}{2})
\gamma^{\mu}(1-\gamma_5)],
\end{equation}
\begin{eqnarray}
A_{Fig.2(c)} &= &i\int \frac{d^4s}{(2\pi )^4}\frac{d^4q}{(2\pi )^4} C_i(q^2) T_r[\chi^{P_1}_{B}(s)\Gamma_{\mu} 
\bar{\chi}^{P_2}_{D^0}(s-q-\frac{P_3}{2})(\not s-\not q+\frac{\not P_1}{2}-\not P_3-m_c) 
 \nonumber  \\
 & ~ &~~~~~~~~~~~~~~~~~~\bar{\chi}^{P_3}_{D^-}(s-q+\frac{P_2}{2})
\gamma^{\mu}(1-\gamma_5)].
\end{eqnarray}
Here $C_i(q^2) $ is a particular Wilson coefficient which may carry a $q^2 $ dependence via the function $G(m_q, \mu, q^2) $ and could be complex. Notice that difference between Figs.2(b) and 2(c) is that the penguin is "time-like" or "space-like", in quark language, respectively. In the B-S formalism, both of these graphs can generate  complex amplitudes as q is an integration variable spanning the entire space.  $\Gamma^{\mu} $ is the appropriate matrix, $ (V\pm A) $, depending on which of the Wilson coefficients is chosen for $C_i(q^2) $.  $\chi_B(s) $ etc. are the phenomenological B-S relativistic wavefunctions which we choose according to Esteve et.al.\cite{E},
\begin{equation}
\chi^P_M(q)=N \gamma_5 (1+\delta \not P) \Phi(P, q),~~~ \bar{\chi}=\gamma_0\chi^+\gamma_0
\end{equation}
where the momentum wavefunction $\Phi(P,q)  $ is, 
\begin{equation}
\Phi(P,q)=exp\{-\frac{1}{2\alpha^2}[ 2(\frac{P\cdot q}{m_M})^2-q^2]\}.
\end{equation}
This is a particularly simple form which in the rest frame of M reduces to

\begin{equation}
\Phi(P,q)\frac{rest frame}{~~} exp\{-\frac{1}{2\alpha^2}[ q_0^2+\vec{q}^2]\},
\end{equation}
Clearly, this form assigns the same strength to the space- and time- oscillators. There is no compelling reason for this. A more general form could be obtained by assigning a different $\alpha $ to $(P\cdot q)^2 $ and $q^2 $ in the choice of $\Phi(P,q) $, but  we don't concern ourselves with such finer details. The values of the parameters $\delta, \alpha$ and N are fixed  according to  Esteve et.al. \cite{E}
\begin{eqnarray}
\delta &=& -\frac{m_a+m_b}{m^2}\\
\alpha &=&\sqrt{-\frac{\pi f_M}{4\sqrt{3}\delta}}\\
N&=&\frac{4\pi}{\alpha^2}
\end{eqnarray}
$m_a$, $m_b$  are  the current masses of quarks  making up   the meson M with the decay constant $f_M$.\\
\begin{flushleft}
{\Large \bf 4. Results}
  
\par
We present our results under the following headings:\\
I. "Effective G-functions".
\end{flushleft}
\par
 We first calculated the "effective G-functions" defined by
\begin{equation}
\bar{G}^{(j)}_q=\frac{<M_1 M_2|O_jG(m_q,\mu,q^2)| B>}{<M_1 M_2|O_j| B>},~~\bar{G}^{(j)}_{q_1q_2}=\frac{<M_1 M_2|O_jG(m_{q_1},\mu, q^2)G(m_{q_2},\mu,q^2)| B>}{<M_1 M_2|O_j| B>},
\end{equation}
where $O_j $ are the different currents. The numerator in eq.(20) is calculated according to eq.(12) where the coefficient $ C_i(q^2) $ for the current product  $O_i $ is replaced by function $G(m_q, \mu, q^2)  $ but the  Lorentz structure of the current product  $O_i $ is kept. The denominator is calculated according to eq.(12) with $C_i(q^2)=1$. The effective G's are independent of j. Thus we drop the superindex henceforth. $\bar{G}_{q_1q_2} $ is relevant to calculations involving a quark-antiquark loop insertion in the gluon propagator. We show the results in Table I where we also contrast the effective G-function with the value calculated in the conventional approach with two chosen values of $q^2 $: $q^2=m^2_b/4 $ and $q^2=m^2_b/2 $.  We notice,  first,  that $\bar{G}_q $ are process dependent and, second, the largest difference between the standard approach and ours is seen for processes involving charm quark in the penguin loop.
\begin{flushleft}
II. CP Asymmetry.
\end{flushleft}
\par
We calculated the CP asymmetry defined as

\begin{equation}
a_{cp}=\frac{\Gamma-\bar{\Gamma}}{\Gamma+\bar{\Gamma}}
\end{equation}
where $\bar{\Gamma} $ represents the rate for the CP conjugate reaction. This calculation was done in three different scenarios:  (a) $ without $ the inclusion of $O(\alpha_s) $ quark-antiquark vacuum 
polarization insertion in the gluon propagator,  (b) $ with $ the $O(\alpha_s) $ vacuum polarization loop in the gluon propagator,  and (c) $ without $  the $O(\alpha_s) $ vacuum polarization loop and $ with~ the~ imaginary~ part~ from~ the~ diagonal~ quark-antiquark~ states $ (for example, $c\bar{c} $ contribution for channel $B^-\rightarrow D^-D^0 $ ) $ removed $,  as motivated by CPT requirement and advocated in \cite{DHP}. Table 2 summarizes the results of these calculations: $B^-\rightarrow K^-K^0 $ decay is not represented in this Table because it is a pure penguin process whereas the CPT cancellation occurs between $Tree\otimes O(\alpha_s^2) Penguin $ and $O(\alpha_s) Penguin \otimes  O(\alpha_s) Penguin $\cite{DHP, GH, LW}.

\begin{flushleft}
III. Comparison with other model calculations.
\end{flushleft}
\par
We  compared the results of calculating the decay rates and CP asymmetries in four different methods as listed below:
\begin{flushleft}
(i) B-S approach:  Here we calculate the decay rates and $a_{cp}$ in our approach but with the gluon vacuum ploariztion loops and the specelike penguin graphs, Fig.2(c), omitted.\\
(ii) As the B-S approach also allows us to calculate the transition form factor $F_0(q^2) $ (defined by Bauer, Stech and Wirbel (BSW) \cite{BSW}), using the wavefunctions (14) and (15)
in a covariant formalism, we attempted a conventional calculation (in the factorization approximation ) with the calculated B-S form factor  $ F_0(q^2) $ and $G_q(q^2) $ evaluated at
 $q^2=m_b^2/4 $ and $m_b^2/2 $.  The B-S formfactor $F_0(q^2) $  can be written down in a form similar to eq.(11-13).  \\
(iii)  A hybrid calculation:  we attempted a conventional calculation ( in the factorization approximation )  with BSW\cite{BSW}  form factors and the effective G function, $\bar{G}_q(q^2) $ of (20) rather than the G-function itself.  \\
(iv)  A conventional calculation (in the factorization approximation) with BSW\cite{BSW}  form factors and  $G_q(q^2) $ evaluated at $q^2=m_b^2/4 $ and $m_b^2/2 $. \\
\end{flushleft}

The results are shown in Table 3. 
\par
We need to point out the following features of Table 3.  First, we found, as was also the case in \cite{SW},  that the absolute branching ratios came out too low by more than two orders of magnitude. Further, the shortfall was decay mode dependent.  The branching ratios appearing in Table 3 under the columns marked Method 1 and 2 were normalized according to the following prescription:
\begin{equation}
B(B^-\rightarrow D^-D^0)=\frac{B(B^-\rightarrow D^-D^0)}{B(\bar{B}^0\rightarrow D^-l^+\nu_l)}\times 1.9\times 10^{-2} 
\end{equation}
\begin{equation}
B(B^-\rightarrow K^-\pi^0, K^-K^0)=\frac{B(B^-\rightarrow K^-\pi^0, K^-K^0)}{B(\bar{B}^0\rightarrow \pi^-l^+\nu_l)}\times 1.8\times 10^{-4} 
\end{equation}
Experimental values of $B(\bar{B}^0\rightarrow D^-l^+\nu_l) $ and  $ B(\bar{B}^0\rightarrow \pi^-l^+\nu_l) $ are from \cite{PDG} and \cite{CLEO}, respectively.  Despite the need to thus normalize the  branching ratios, the CP asymmetry ought to be reliable.
 Second, in our  B-S approach,  q is integrated over, but in methods where q has to be chosen (Methods 2 and 4), $a_{cp} $ depends sensitively  on the choice of $q^2 $. This is due to the fact that the imaginary 
part of the $c\bar{c} $ penguin loop depends strongly on the choice of $q^2 $ in the range $m_b^2/4\leq q^2 \leq m_b^2/2 $.

\par
\begin{flushleft}
{\Large\bf 5. Discussion}
\end{flushleft}
\par
We begin by listing the parameter set we have used in our calculations.
\begin{eqnarray}
~& m_u=m_d=5MeV,~~ m_s=160MeV, ~~m_c=1.35GeV,~~m_b=5.0GeV. &~\nonumber \\
~ & f_B=f_D=200MeV,~~f_K ~~and~~ f_{\pi}~~ from \cite{PDG}    &~ \nonumber  \\
~ & \alpha_B=0.71GeV,~~ \alpha_D=0.41GeV, ~~\alpha_K=0.33GeV, ~~\alpha_{\pi}=0.33GeV  & ~\nonumber \\
~& CKM~ angles: A=0.90, ~~\lambda=0.22,~~ \rho=-0.12,~~\eta=0.34 \cite{AL}  & ~
\end{eqnarray}
In calculating errors in the CP asymmetry $a_{CP} $ , we allowed a $\pm 10\% $ error in the above listed $\alpha_B $,  $\alpha_D $ and $\alpha_K $. To put this in perspective in terms of the 
variation of parameters in (17)-(19),  allowing $m_b$ to vary in the range (4.8-5.2) GeV and $f_B $ in the range (180-220)MeV, $\alpha_B $ changes by $\pm 7\% $ around its central value; similarly,  allowing $m_c $ to vary in the range (1.2-1.6) GeV and $f_D $ in the range (180-220)MeV, $\alpha_D $ changes by $\pm 12\% $ around its central value;  and allowing $m_s $ vary in the range (150-200) MeV changes $\alpha_K $ by $\pm7\% $. \\

We have done a calculation of CP asymmetries  in a selected few  hadronic two-body B decays in a Bethe-Salpeter formalism and contrasted the results with  those  of the conventional factorized approach. We have  also investigated the effect of including the gluon quark-antiquark vacuum polarization loops. We find that $a_{cp}$ depends strongly on the method adopted to calculate it. In particular, for decays into charmless mesons, the conventional factorization approach results in an asymmetry which depends very strongly on the choice of $q^2 $ (see Methods 2 and 4 of Table 3).  This is due to the fact that $c\bar{c} $ loops generate an imaginary part which is sensitive to the choice of $q^2$ in the range $m_b^2/4\leq q^2 \leq m_b^2/2 $. Our method, on the contrary,  smears over $q^2$ through a convolution of meson wavefunctions. \\

A shortcoming of our approach (which we  share with \cite{SW}) is that the branching  ratios are too low. A single overall normalization, independent of the decay mode, does not correct the situation.  Thus, we normalized  $B(B^-\rightarrow D^-D^0) $ to the measured branching ratio for semileptonic  $ \bar{B}^0\rightarrow D^-l^+\nu_l  $, and $B(B^-\rightarrow K^-\pi^0) $ and $ B(B^-\rightarrow  K^-K^0 ) $ to the semileptonic $ \bar{B}^0\rightarrow \pi^-l^+\nu_l $ branching ratio.   However, $a_{cp}$ ought to be reliable. In technical terms the low value of the decay rates is related to the time-oscillators which further dampen the contribution of the space-oscillators in our covariant wavefunctions. Perhaps the use of more sophisticated bound state wavefunctions\cite{ TZ} would alleviate this difficulty.

\newpage
\begin{flushleft}
{\Large\bf Figure Captions} \\

Fig.1(a): $O(\alpha_s) $ penguin graph for $b\rightarrow s q\bar{q} $ transition.\\
Fig.1(b): Penguin graph for $b\rightarrow s q\bar{q} $ transition with $O(\alpha_s) $ gluon quark-antiquark vacuum polarization loop. \\
Fig.2(a):   Tree graph arising from operators $O_1$ and $O_2$ leading to the amplitude in eq.(11). \\
Fig.2(b):   Graph arising from operators $O_3-O_{10}$ leading to the amplitude in eq.(12). \\
Fig.2(c):   Graph arising from operators $O_3-O_{10}$ leading to the amplitude in eq.(13). \\
\end{flushleft}

\newpage
\begin{small}
\begin{table}
\begin{center}
\caption{The effective G functions of Penguin  diagrams for some processes. See Eq.(20 ) for definitions. Last two rows list the values of $G_q$ at $q^2=m_b^2/4 $ and $q^2=m_b^2/2 $.}

\begin{tabular}{|c|c|c|c|c|c|c|c|c|}
\hline
&&&&&&&& \\
Channel      & $\bar{G}_u  $        &   $\bar{G}_s  $  & $\bar{G}_c  $ & $\bar{G}_{uu}  $ &$\bar{G}_{us}  $ & $\bar{G}_{uc}  $ & $\bar{G}_{cs}  $ & $\bar{G}_{cc}  $ \\ \hline \hline 
$B^-\rightarrow D^-D^0 $ &  -6.7+1.8i & -2.1+1.8i & 0.24+0.32i & 42-25i & 11-16i & -4.1-1.4i & -1.9+0.10i & -0.095+0.57i  \\ \hline
$B^-\rightarrow K^-\pi^0 $ & -7.1+1.4i & -2.5+1.4i & 0.19+0.67i & 49-21i & 16-14i & -2.7-4.6i & -1.9-1.5i & -0.81+0.41i  \\ \hline
$B^-\rightarrow K^-K^0 $ &  -7.1+1.4i & -2.5+1.4i & 0.19+0.68i & 49-21i  & 16-14i & -2.8-4.7i &  
 -1.9-1.5i & -0.82+0.41i \\ \hline
$ q^2=m_b^2/4 $ &  \multicolumn{8}{|l|}{ $ G_u=-7.2+2.1i~~~~G_s=-2.5+2.1i~~ ~~G_c= 0.87$}  \\ 
$ q^2=m_b^2/2 $ &  \multicolumn{8}{|l|}{ $ G_u=-7.6+2.1i~~~~G_s=-3.0+2.1i~~~~ G_c=0.65+1.7i $}  \\  \hline

\end{tabular}
\end{center}
\end{table}
\end{small}

\begin{small}
\begin{table}
\begin{center}
\caption{CP asymmetries $a_{cp} (\%)$ under different scenarios: column 2, without gluon vacuum ploarization loops; column 3, with quark-antiquark gluon vacuum polarization loops; column 4, without gluon vacuum polarization loops and  imaginary parts arising from diagonal channels removed. Figures in  parentheses correspond to a calculation where  spacelike penguins (Fig.2(c)) were ignored. }
\begin{tabular}{|c|c|c|c|}
\hline
&& &\\
Channel  &   gluon vacuum  &  gluon vacuum &   imaginary part from      \\
& ploarization excluded & polarization included & diagonal channels removed \\ \hline \hline
$B^-\rightarrow D^-D^0 $ & 2.7 (2.9)&  2.1 (2.2) & 2.7 (2.9)  \\
$B^-\rightarrow K^-\pi^0 $ & 4.0 (2.9) & 3.1 (2.3) &  3.4 (2.4)   \\ \hline
\end{tabular}
\end{center}
\end{table}
\end{small}

\newpage
\begin{small}
\begin{table}
\begin{center}
\caption{The calculated CP asymmetries and branching ratios in different methods
Method 1: $O(\alpha_s) $ calculation in our B-S approach; Method 2: Factorization approximation with B-S form factors. First entry uses $q^2=\frac{m_b^2}{4}$ and second $q^2=\frac{m_b^2}{2}$; Method 3: Factorization approximation with BSW from factors and effective $\bar{G}_q$; Method 4: Factorization approximation with BSW form factors. First entry uses $q^2=\frac{m_b^2}{4}$ and second $q^2=\frac{m_b^2}{2}$.  }

\begin{tabular}{|c|c|c|c|c|}
\hline
&&&& \\
 Channel  & Method 1 & Method 2 & Method 3  & Method 4\\
                   &  Br and $a_{cp}(\%) $ & Br and $a_{cp}(\%) $ &Br and $a_{cp} (\%)$ &Br and $a_{cp}(\%) $ \\ \hline\hline
$B^-\rightarrow D^-D^0 $ &  $5.3\times 10^{-4} $, ~~$2.9\pm 0.1 $  & $ 5.2\times 10^{-4} $, ~~3.3  & $4.7\times 10^{-4} $, ~~2.8 & $ 4.6\times 10^{-4} $, ~~3.2\\
                                               & & $ 5.3\times 10^{-4} $, ~~3.6  &  & $4.7\times 10^{-4} $,~~3.5 \\ \hline
$B^-\rightarrow K^-\pi^0 $  & $ 7.0\times 10^{-6}$,~~ $2.9\pm 0.2$ &  $7.8\times 10^{-6}$, ~~ 0.68 &  $1.1\times 10^{-5}$, ~~ 4.6 & $1.2\times 10^{-5}$, ~~0.91  \\
                                                 & & $8.0\times 10^{-6}$, ~~ 6.2 &  & $1.3\times 10^{-5} $,~~9.4 \\ \hline
$B^-\rightarrow K^-K^0 $ & $4.0\times 10^{-6}$,~~$-7.1\pm 0.6 $ & $4.3\times 10^{-6}$,~~ -19 & $1.0\times 10^{-6}$, ~~ -7.0 &  $1.1\times 10^{-6}$, ~~ -19  \\  
					 &  & $5.1\times 10^{-6}$, ~~ -5.3  & & $1.2\times 10^{-6}$, ~~ -5.3  \\ \hline	
\end{tabular}
\end{center}
\end{table}
\end{small}

\end{document}